# Applying Resonant Spin Flippers with Poleshoes and Longitudinal Radio Frequency Fields to Time of Flight MIEZE

**N.Geerits[1], S.R.Parnell[1], M.A.Thijs[1], W.G.Bouwman[1] and J.Plomp[1]**

[1]Faculty of Applied Sciences, Delft University of Technology, 2629JB Delft, The Netherlands

**Abstract**. A time of flight MIEZE spectrometer study is presented. The instrument uses solenoid radio frequency (RF) spin flippers with square pole shoes and a magnetic yoke. These flippers can achieve higher static fields than conventional resonant RF spin flippers, which employ an air core. High fields are crucial for the construction of a high resolution and compact MIEZE spectrometer. Using both types of flippers two MIEZE spectrometer configurations are constructed and compared on the same beam line. It was demonstrated that the pole shoe/solenoid coil RF flippers can achieve a MIEZE signal, which is similar in quality to the conventional reference setup. The highest obtained modulation frequency was 100 kHz.

## 1. Introduction

Modulated IntEnsity by Zero Effort (MIEZE) is a polarised neutron spin echo technique, used to measure quasi-elastic scattering [1-5]. As the name suggests, a MIEZE instrument will modulate the neutron beam intensity in time. In the absence of a sample the amplitude of these intensity oscillations are maximal and indicate the resolution function of the instrument. When a sample with the appropriate dynamics (~0.03-10 μeV) is placed in the instrument this amplitude reduces, this is a measure for the energy transfer between the sample and neutron. MIEZE instruments employ two Neutron Resonant Spin Echo coils (NRSE coils), which induce a longitudinal Stern-Gerlach effect. The first NRSE coil causes the two neutron spin states to go out of phase, while the second coil overcompensates the effects of the first coil. As a result the phase difference between the two states will diminish as the neutrons approach the focus point where they are measured (figure 1, adapted from reference 3). This allows one to place the sample after all the spin manipulation components, including the analyser. When an inelastic scattering event takes place between the last NRSE coil and the detector the two spin states will arrive at the detector slightly out of phase and the amplitude of the intensity modulation will decrease. The maximal achievable phase difference between the two spin states at the sample determines the resolution of the spectrometer known as the spin echo time [6]. Typically MIEZE spectrometers utilize rectangular air core electromagnets with RF coils (resonant flippers). The RF frequency is tuned such that $\omega = \gamma B$. The MIEZE modulation frequency is twice the difference frequency between the two NRSE coils [4]. The flippers in this study employ square pole shoes, which generate a static magnetic field perpendicular to the beam, and a solenoid RF coil, which creates an RF field parallel to the beam. Due to the yoke and pole shoes these flippers can achieve higher fields than resonant flippers [7]. The RF fields are modulated using a *1/t* function; this ensures that the flippers flip the entire bandwidth of a pulsed beam [8].

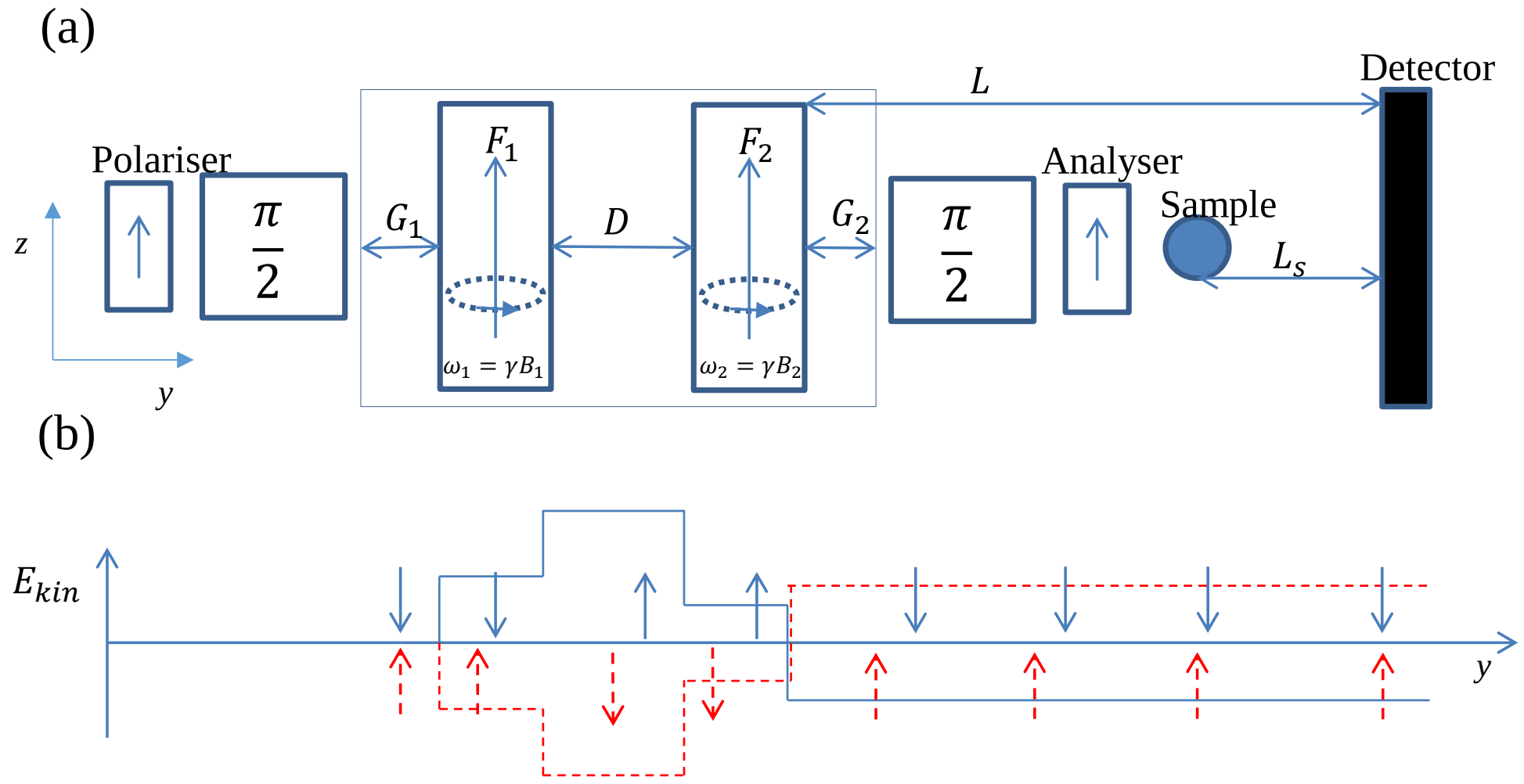

**Figure 1.** (a) Schematic of a MIEZE spectrometer as implemented in the RF flipper with poleshoe setup. The flippers $F_1$ and $F_2$ are contained within a guide field and generate a static field in the $\hat{z}$ direction and an RF field in the $\hat{y}$ direction. The $\pi/2$ flippers project the neutron spin on the x-axis. (b) The kinetic energy of the neutron wavefunction for each spin state (denoted by the blue solid line/arrow and red dashed line/arrow) as a function of position along the beam. The distance between the arrows denotes the phase difference between the two neutron spin states at a certain position in the instrument.

## 2. Theory

### *2.1. Radio Frequency Spin Flippers*

The core of every MIEZE setup to date is a set of two (or more) spin flippers. This study focuses on a type of flipper that utilizes square pole shoes to generate the static magnetic field. The main advantage of these pole shoes is they allow the generation of high homogeneous magnetic fields in combination with relatively low RF coil current without having material in the beam. The static magnetic field is orientated perpendicular to the neutron flight path, while the RF field is parallel to the flight path. Therefore the total field is given by

$$\vec{B} = B_i\hat{z} + B_{rf}\cos(\gamma B_i t + \phi)\,\hat{y} \tag{1}$$

Where the index *i* denotes the RF flipper (one or two). The linearly oscillating RF field can be split into two rotating fields, which rotate in opposite directions. If $B_i$ is large compared to $B_{rf}$ one can ignore the field which rotates in the direction opposite to the Larmor precession [9]. Therefore the total field can be approximated by

$$\vec{B} = B_i\,\hat{z} + B_{rf}[\cos(\gamma B_i t + \phi)\,\hat{x} - \sin(\gamma B_i t + \phi)\,\,\hat{y}] \tag{2}$$

The Schrödinger equation has been solved for a neutron in this magnetic field in [10,11]. It turns out that for

$$B_{rf} = \frac{\pi v}{\gamma d} \tag{3}$$

with *v* the neutron velocity and *d* the flipper length, the flipper induces a maximal energy difference between the spin up and spin down states of the neutron. For a pulsed beam $v = v(t)$, thus equation (3) can be written as followed:

$$B_{rf} = \frac{\pi S}{\gamma d t} \tag{4}$$

with *S* the distance between the flipper and the pulsed source [6].

*2.2. MIEZE Focusing Condition*

To prevent depolarisation of the beam the neutrons must either traverse a zero field chamber or a guide field. Traditional MIEZE spectrometers opt for a zero field chamber; however the poleshoe magnets, due to their larger dimensions, are placed inside of a guide field. The guide field prevents depolarisation of the beam, however it also introduces additional precession, which must be accounted for when determining the MIEZE focus. In the case of a zero field chamber the MIEZE focus is given by [4]:

$$L = \frac{D}{\frac{\omega_2}{\omega_1} - 1} + \frac{d}{2} \quad (5)$$

here $D$ is the distance between the two flippers, d the length of a flipper and $\omega_i = \gamma B_i$ the frequency of the i-th flipper. If a guide field around both flippers is introduced the focusing condition becomes:

$$L = \frac{\Delta\omega d - (2\Delta\omega - \omega_g)G_2 + (2\omega_1 - \omega_g)D + \omega_g G_1}{2\Delta\omega} + G_2 \quad (6)$$

with $\Delta\omega = \omega_2 - \omega_1$, $\omega_g$ the Larmor frequency of the guide field and $G_1$ and $G_2$ are given in figure 1. When this condition is satisfied the expectation value for the spin in the x-direction is given by [4]:

$$< \sigma_x >= \cos(2\Delta\omega t) \quad (7)$$

But if the detector is out of focus by a small amount $\Delta L$, the expectation value becomes:

$$< \sigma_x >= \cos(2\Delta\omega[t - \frac{m\lambda\Delta L}{h}]) \quad (8)$$

this would be acceptable for a perfectly monochromatic beam, however for a small wavelength spread, $\Delta\lambda$, averaging over the bandwidth is required. Assuming a Gaussian wavelength distribution the out of focus MIEZE signal is given by:

$$< \sigma_x >= \int_{-\infty}^{\infty} \frac{\mathrm{e}^{-\frac{(\lambda-\lambda_0)^2}{2\Delta\lambda^2}}}{\sqrt{2\pi\Delta\lambda^2}} \cos\left(2\Delta\omega\left[t - \frac{m\lambda\Delta L}{h}\right]\right) d\lambda = e^{-2\left(\frac{\Delta\omega m\Delta L\Delta\lambda}{h}\right)^2} \cos(2\Delta\omega\left[t - \frac{m\Delta L\lambda_0}{h}\right]) \quad (9)$$

For a pulsed beam $\lambda_0$ is time dependent, meaning the MIEZE frequency will shift if the detector is out of focus.

$$< \sigma_x >= e^{-2\left(\frac{\Delta\omega m\Delta L\Delta\lambda}{h}\right)^2} \cos(2\Delta\omega\left[1 - \frac{\Delta L}{L}\right] t) \quad (10)$$

This frequency shift is a useful tool for finding the exact MIEZE focus experimentally. In addition placing the detector out of focus, or using a detector of non-zero thickness will result in an amplitude drop [12]. One can see that the focal spot size is dependent on the wavelength resolution. In the described experiment a chopper was used with a wavelength resolution of 4.7%, meaning the focusing condition (equation (5)) must be satisfied with high accuracy to prevent loss of signal.

## 3. Method

The goal is to demonstrate that the solenoid coil RF flippers, figure 2 (b), are competitive with conventional resonant flippers figure 2 (a). Solenoid coil RF flippers are different in that they use magnetic poleshoes and a yoke to generate a large static magnetic field at relatively low coil currents, removing the need for water cooling, furthermore the RF field is generated parallel to the beam, rather than perpendicular to it as is the case with conventional RF flippers, using a solenoid. A drawback to this technique, is that the field regions are less sharply defined, compared to conventional RF flippers. However this should not produce any problems, if the envelope of the RF field does not protrude, beyond the poleshoes, where the resonance condition is not met. Nonetheless this drawback may lead to the question of how applicable these spin flippers are to MIEZE. To answer this question two setups were constructed: first, a reference MIEZE spectrometer using two bootstrapped [4,13] resonant flippers contained in a zero field chamber, originally from the 2010 version of the FLEXX instrument at HZB [14], and a second setup using two flippers, with poleshoes, contained inside a guide field. Finally the MIEZE signals from both setups were compared. Note that we measure the visibility of the signals and not the polarisation. The difference being that the visibility is the measured polarisation, which can differ from the actual polarisation, due to averaging effects.

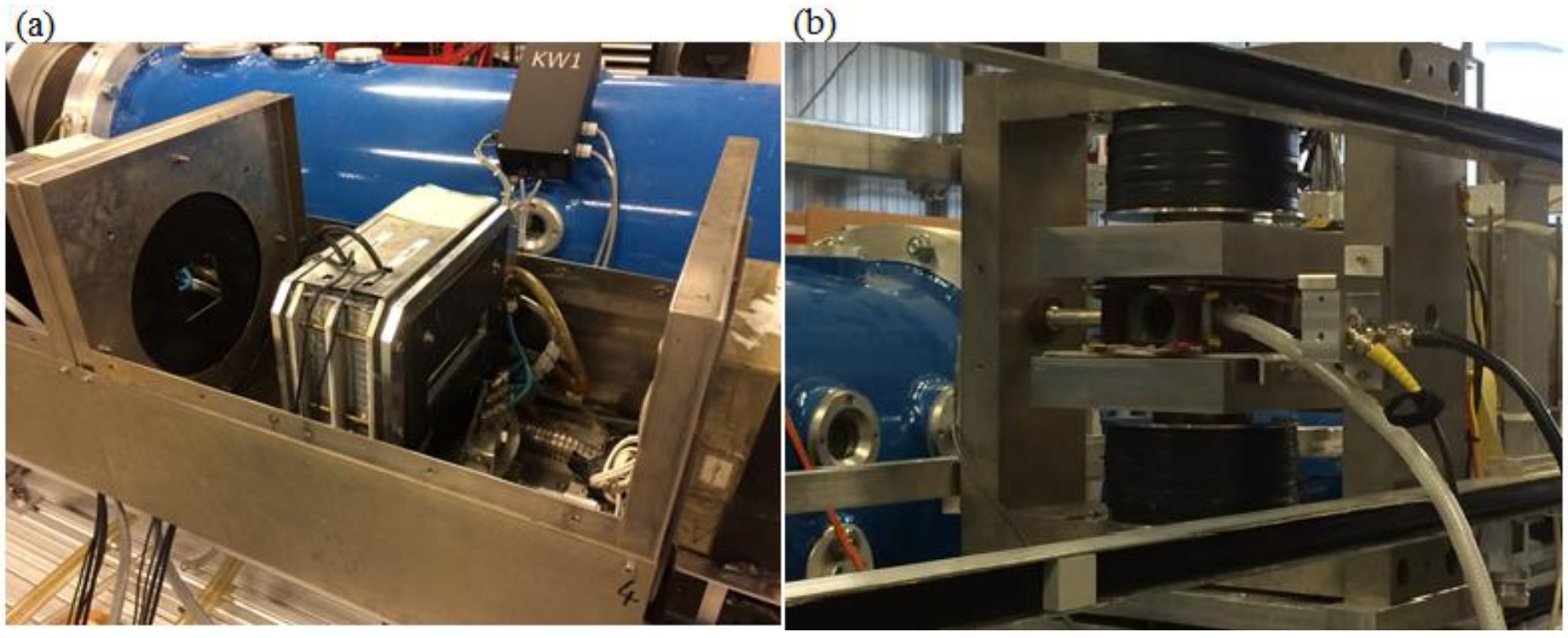

**Figure 2**. (a) Photograph of a bootstrapped (N=2) resonant flipper. For more information on these types of coils see [14]. (b) Photograph of an RF flipper using a magnetic yoke and poleshoes. Due to these magnetic poleshoes large static magnetic fields can be generated at relatively low coil currents.

### *3.1. Reference Setup*

The reference setup (figure 3) is constructed using a double disk chopper, providing an effective pulse frequency of 50Hz. The chopper is followed by a polariser and a slit (1cm x 1cm). Next is an RF flipper (60 kHz) which allows measurement of both spin states, allowing determination of the beam polarisation. Further downstream the core of the setup is situated: a $\pi/2$ rotation element [15], zero field chamber containing both resonant flippers and a second $\pi/2$ rotation element. Finally at the end of the setup is a slit set (1cm x 1cm), analyser and a ${}^{3}_{2}He$ detector.

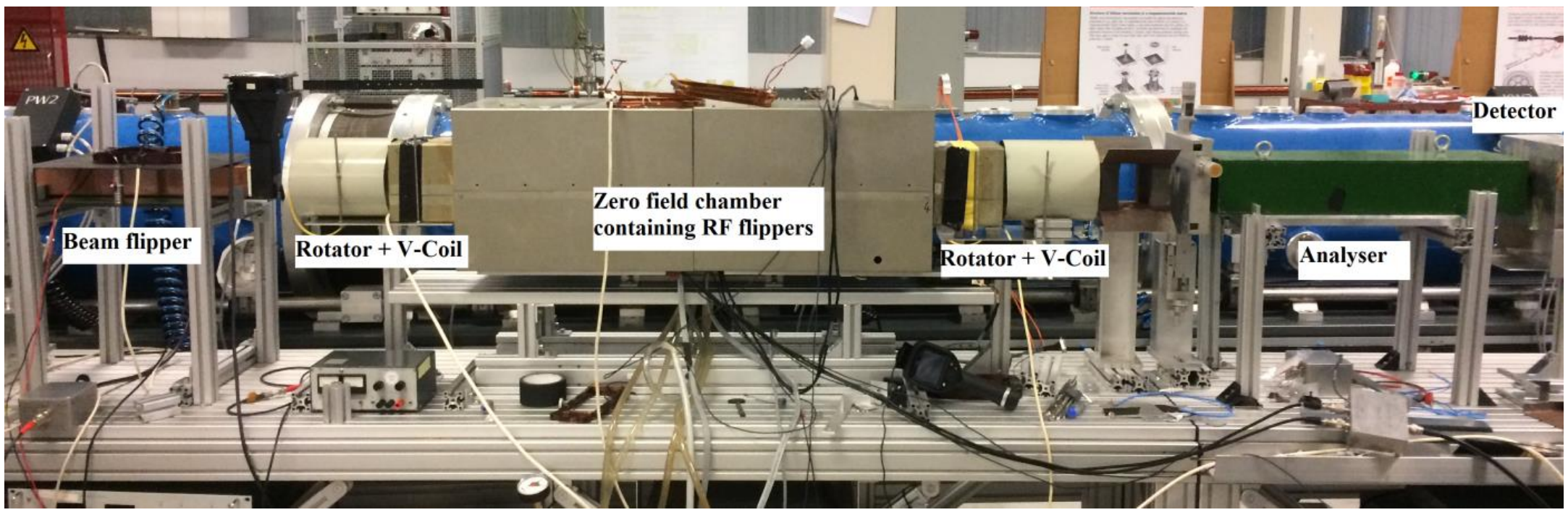

**Figure 3**. Photograph of the reference setup: from left to right one can see the beam flipper, $\pi/2$ rotation element, the zero field chamber containing the two resonant flippers, $\pi/2$ rotation element, analyser and detector.

As was demonstrated in section 2.2, the focal spot size is very small in this setup. To remedy this, an improvised PSD consisting out of 16 1cm diameter ${}^{3}_{2}He$ tubes placed behind each other is used. The PSD is placed into the beam under a slight angle (~3°), so that each tube is able to "see" a piece of the beam as illustrated in figure 4 (a). This allows scanning of a large focal area at once. This PSD also allows mapping of the focal spot, enabling us to compare with our prediction from section 2.2 is.
The resonant flippers were operated at frequencies of 181.6 kHz and 156.6 kHz. Since each RF flipper is a bootstrapped pair, the effective zero field precession frequency is twice the flipper frequency [4, 13] (i.e. 363.2 kHz and 313.2 kHz respectively). This results in an expected modulation frequency of 100 kHz. The RF signal was generated using a function generator and RF amplifier. A custom function, consisting of a cosine multiplied by a 1/t function (given by equation 4), was generated using Matlab, figure 4 (b), and outputted using the frequency generator. The flippers were placed 0.675 m apart, resulting in a calculated focal point at 4.230 m from the last flipper.

(a) (b)

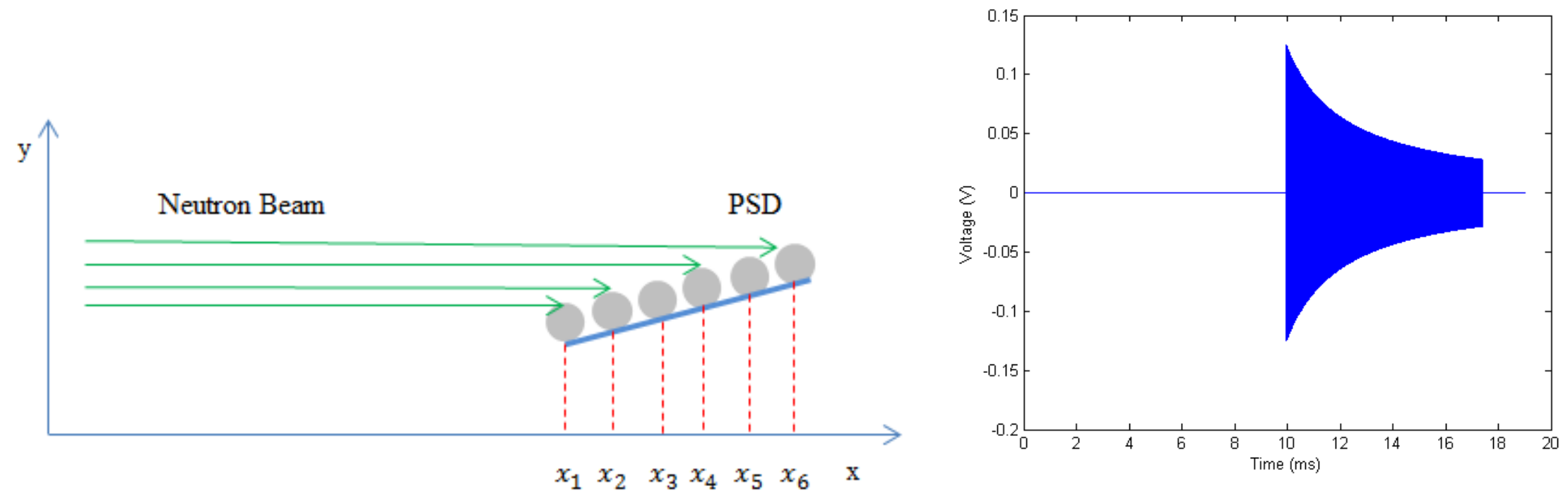

**Figure 4**. (a) Schematic of improvised PSD placed at a small angle to the beam. Each tube "sees" the neutron beam at a different location denoted by $x_i$. (b) Function used to drive RF coils. The coils are turned on at 1.8 Å and off at 8 Å.

### 3.2. *Experimental MIEZE Setup*

The experimental MIEZE setup (figure 5), is similar to the reference setup. The zero field chamber containing the resonant flippers was replaced with a 2 m long guide field containing two flippers with poleshoes. In addition the PSD was exchanged for a single 1 cm $^3_2He$ tube. To increase the focal spot size the two chopper disks were moved closer together.

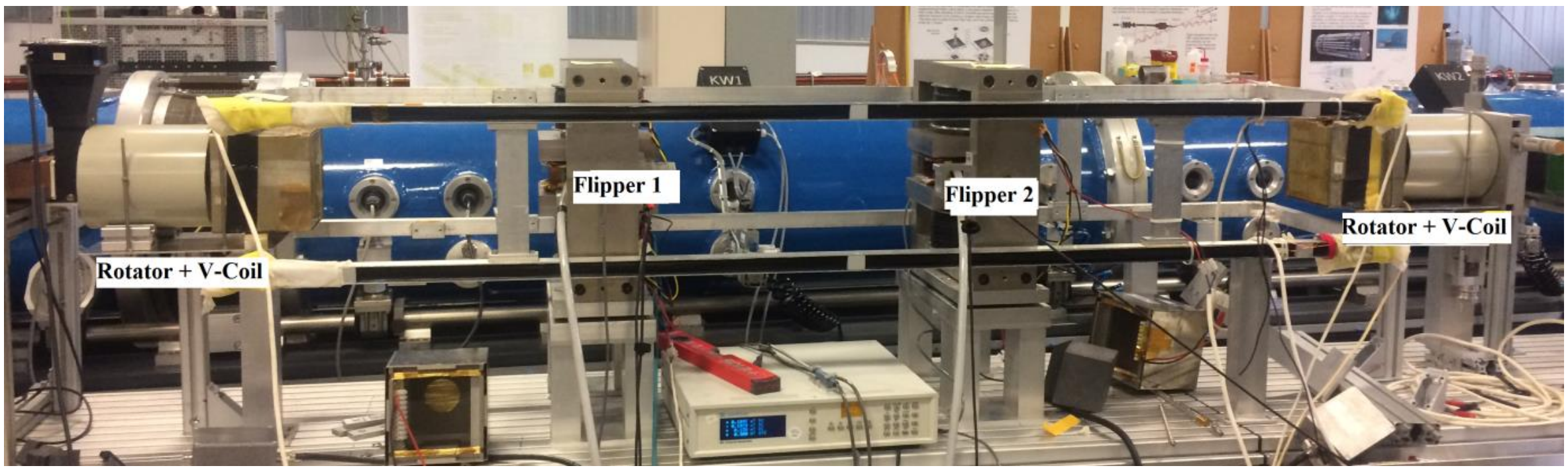

**Figure 5**. Photograph of the solenoid coil RF MIEZE setup: from left to right - $\pi$ flipper, $\pi/2$ rotation element and the start of the guide field, two solenoid coil RF flippers, second $\pi/2$ rotation element and the end of the guide field (analyser and detector not shown but same as shown in figure 3).

Three different flipper frequencies were used, in increasing order, to determine the focal point, as at lower modulation frequencies the focal spot is broader. Again, the current through the RF coils was modulated by a 1/t function. The experimental parameters are shown below in table 1. Note that the ratio between the two flipper frequencies remains constant as they are increased. At a 100 kHz modulation frequency the calculated focal point was found to be 4.507 m from the last flipper.

| $f_1$ | $f_2$ | $2\Delta f$ | $G_1$ | $G_2$ | $D$ | $B_g$ |
|---|---|---|---|---|---|---|
| 61.49 kHz | 71.49 kHz | 20 kHz | 0.383 m | 0.614 m | 0.720 m | 0.6 mT |
| 122.98 kHz | 142.98 kHz | 40 kHz | 0.383 m | 0.614 m | 0.720 m | 0.6 mT |
| 307.45 kHz | 357.45 kHz | 100 kHz | 0.383 m | 0.614 m | 0.720 m | 0.6 mT |

**Table 1**. Table containing parameters for the experiments conducted using the square poleshoe flippers. The various parameters and physical meaning can be found in figure 1 and section 2.2.

## 4. Results

During these experiments the polarisation was 93% at 2 Å, 86% at 3 Å and 78% at 4 Å in non-precessing mode. Furthermore the flipping efficiency of the bootstrapped flippers in the reference setup was 94%, while the flipping efficiency of the poleshoe flippers was 95%. In figure 6 a comparison of the MIEZE signal from the conventional bootstrap coil setup and the signal from the solenoid coil flipper setup can be seen in both the time and frequency domain. The frequency domain picture was obtained using a fast Fourier transform (FFT) in Matlab.

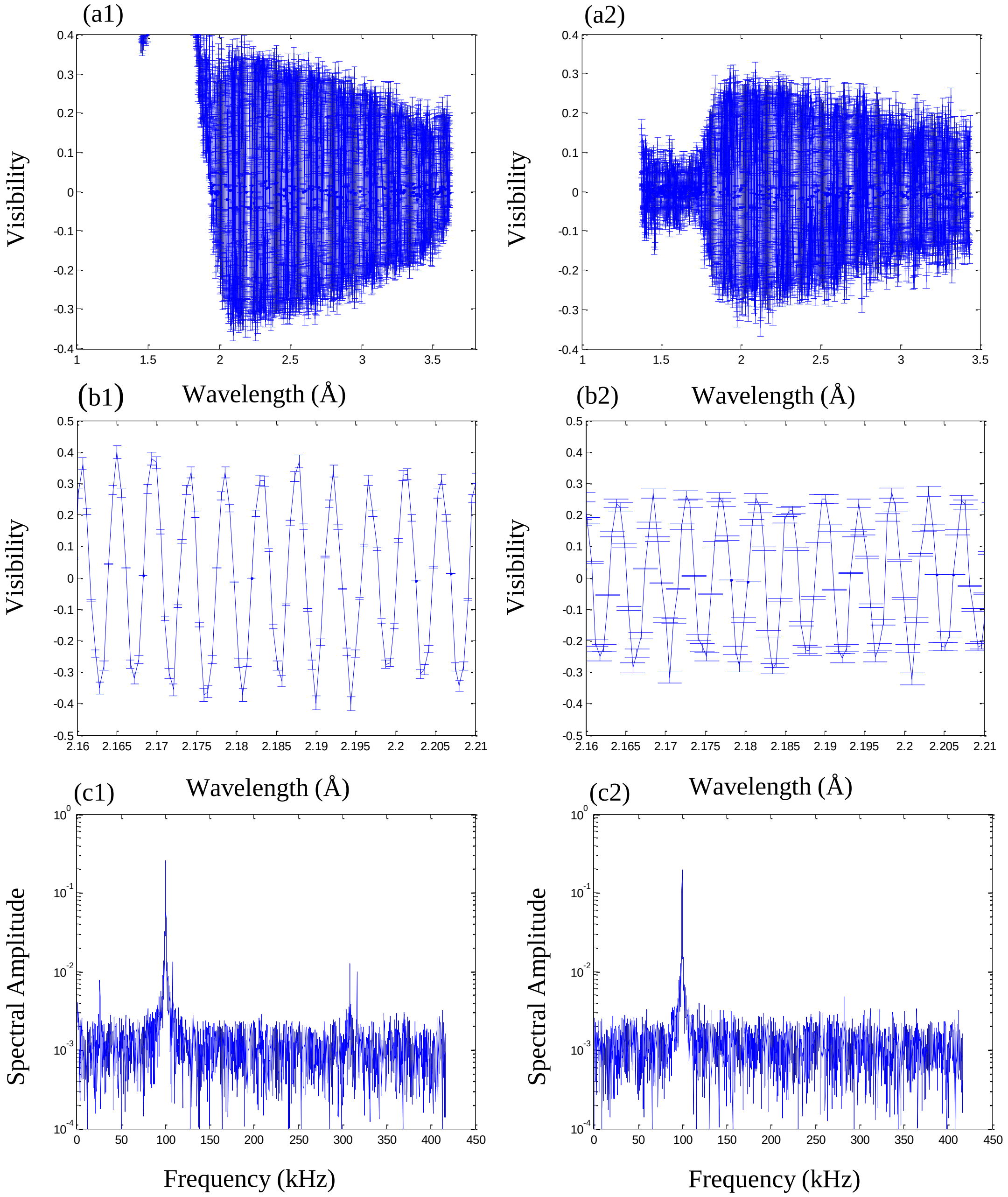

**Figure 6.** Signals from the reference setup (a1-c1) and the poleshoe/solenoid coil RF flipper MIEZE setup (a1-c2). Figures a1 and a2 show the full wavelength range, however the modulation is obscured due to the high frequency, hence in b1 and b2 we show a portion of the signal and in c1 and c2 FFTs of the entire signal can be seen. Note the declining envelope in a1 and a2, which is caused by a combination of factors such as the increased sensitivity to detector thickness and defocusing at higher wavelength and the overall declining polarisation of the analyser at higher wavelengths

In figure 7 the MIEZE signal from the experimental MIEZE setup are shown at 20, 40 and 100 kHz.

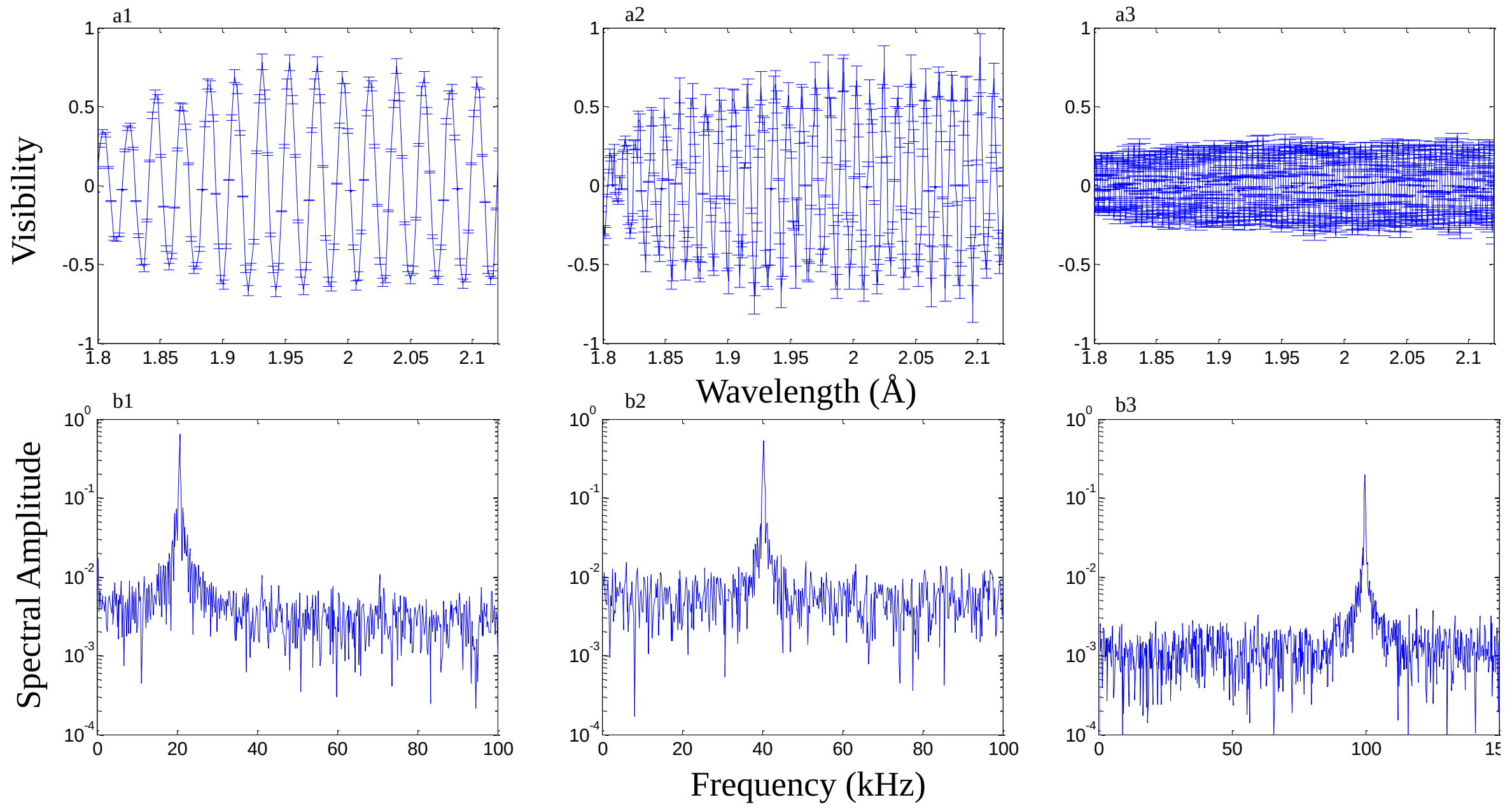

**Figure 7**. MIEZE signal (a1-a3) and FFT (b1-b3) for 20 kHz, (a1-b1), 40 kHz (a2-b2) and 100 kHz (a3-b3). Note the declining amplitude as the frequency increases. This is due to the fact that the detector thickness limits the time resolution. Furthermore errors due to timing inaccuracies (i.e. triggering of the function generators), become prominent as the frequency increases.

In figure 8 (a) the MIEZE amplitude as a function of displacement from the true focus is shown. Finally in figure 8 (b) the frequency shift as a function of detector displacement is shown, using an FFT. Both of these data sets were obtained using the reference MIEZE setup, which employed the bootstrapped RF flippers.

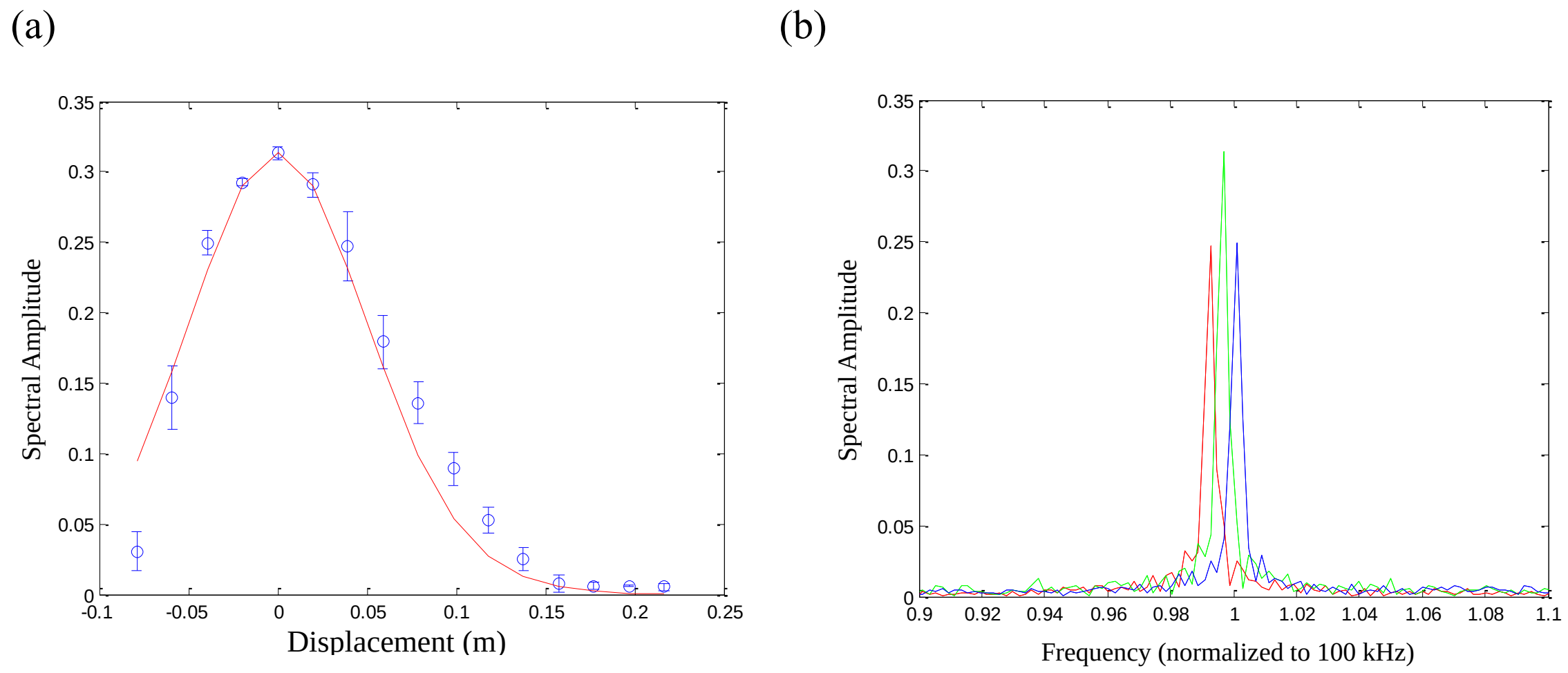

**Figure 8**. (a) Spectral amplitude of the MIEZE signal as a function of the displacement from the true focus, measured with a PSD. The red solid line shows the model (eq. 10, normalized to data), while the blue points represent the measured data. (b) FFT of the MIEZE signal for various positions on the PSD. The central peak at 100 kHz is the in-focus signal. The frequency shift observed in the blue and red curve is due to a $\pm$2cm displacement from the focal point. The expected frequency shift according to eq. 8 is $\omega' - 2\Delta\omega \cong \pm 475$Hz. The observed frequency shift is $\pm$400Hz, with an uncertainty of 50Hz (pulse frequency).

## 5. Discussion and Outlook

With the results shown in figure 6 and figure 7 we have demonstrated that the poleshoe/solenoid coil flippers are capable of creating a MIEZE signal with a visibility similar to the reference MIEZE setup. Figure 7 demonstrates a decrease in visibility as the MIEZE frequency is ramped up, this to be due to the thickness of the detector, which determines its time resolution $\Delta t = d/v = 5\mu s$ (for 1 cm thick detector and 2 Å neutrons and assuming uniform absorption across the detector). This is equal to the Nyquist criterion for a 100 kHz signal, thus in future experiments with larger modulation frequencies a thin detector, such as the CASCADE type [16] is needed to avoid this limitation. Additionally visibility is reduced at higher frequencies due to timing accuracy of the chopper, which is used to trigger the RF flippers. The standard error was determined to be 0.8 μs, by measuring the jitter in the chopper rotation frequency, which causes the visibility to drop to 60% of its base value. Figure 8 (a) shows the focal spot of the MIEZE setup, which is in good agreement with the theory. Note that only a very small displacement is needed to de-focus the MIEZE signal altogether, this also demonstrates the need for a thin detector and/or a good wavelength resolution [12]. Figure 8 (b) shows the frequency shift resulting from placing the detector out of focus. This is a phenomenon unique to time of flight MIEZE. The frequency shifts appear to be in good agreement with the theory, after taking the position uncertainty into account.

The aim of this study is to pre-commission a MIEZE mode on the Larmor instrument at the ISIS pulsed neutron and muon source. The flippers at Larmor are capable of reaching frequencies of up to 3 MHz, due to the poleshoe design. Spin echo times of up to 20 ns (for 10 Å neutrons) at a modulation frequency of 1 MHz could be reached. The MIEZE mode on Larmor could be useful for studying dynamics of magnetic systems, quasi-elastic neutron scattering and possibly MISANS [5].

## 6. Conclusion

We have demonstrated that poleshoe/solenoid coil flippers are capable of producing MIEZE signals of similar quality compared to the reference MIEZE setup, with modulation frequencies of up to 100 kHz. Hence a MIEZE setup on the Larmor instrument at the ISIS pulsed neutron and muon source can be implemented, reaching spin echo times of up to 20 ns. However due to the high modulation frequency a thin detector, such as the CASCADE type with high time resolution is required.

## Acknowledgements

We would like to thank the TU-Delft Reactor Institute Delft for access to beam time and the Helmholtz Center Berlin and Dr Klaus Habicht for providing the bootstrapped resonant flippers and zero-field chamber. This research was funded by a NWO groot grant No. LARMOR 721.012.102.